\documentclass{article}
\usepackage{epsfig,amsmath,amssymb}
\newcommand{\bfr}{\begin{flushright}}
\newcommand{\efr}{\end{flushright}}

\def\Nc{\mathcal{N}}
\def\ds{\displaystyle}
\def\bea{\begin{array}{c}}
\def\ea{\end{array}}
\def\be{\begin{equation}\bea\ds}
\def\ee{\ea\end{equation}}
\def\bee{\begin{equation}\begin{array}{rcl}\ds}
\def\eee{\end{array}\end{equation}}

\def\d{{\rm d}}

\def\Tr{\text{Tr}\,}

\newcommand\Trule{\rule{0pt}{2.6ex}}
\newcommand\Brule{\rule[-1.2ex]{0pt}{0pt}}

\begin{document}
\title{Calculation of glueball spectra in supersymmetric theories via holography
\thanks{Presented at  the Low x workshop, May 30 - June 4 2013, Rehovot and
Eilat, Israel}%
}
\author{Ivan Gordeli and Dmitry Melnikov\footnote{gordeli@iip.ufrn.br, dmitry@iip.ufrn.br}\footnote{preprint ITEP-TH-33/13}
\\
{\small
\parbox{\linewidth}{
\begin{center} International Institute of Physics, UFRN, \\
Av. Odilon G. de Lima, 1722, Natal 59078-400, RN, Brazil \\
 \& \\
 Institute for Theoretical and Experimental Physics \\
 B.~Cheremushkinskaya 25, Moscow 117218, Russia
 \end{center}}
}
 \smallskip\\
}

\maketitle
\begin{abstract}
Lattice simulations currently present the only way to access non-per\-tur\-ba\-ti\-ve data in strongly coupled theories from a first principle calculation. However, in supersymmetric theories this valuable tool is not available due to the technical \emph{sign problem}. We are going to demonstrate that in the case of glueball spectra a good quantitative estimate for the lightest states of low spin can be obtained by means of the \emph{holographic approach}. We will review the results of the calculation in the singlet glueball sector of the $\mathcal{N}=1$ supersymmetric Klebanov-Strassler model. We come up with a prediction of the spectrum of lightest glueballs in (large $N_c$) $\Nc=1$ supersymmetric Yang-Mills theory.
\\
~
\\
PACS number(s): 04.40.Nr, 04.70.Bw, 11.27.+d
\end{abstract}

\section{Introduction}

Although QCD glueballs have not been identified unambiguously from experiment, their spectrum is believed to be known from the lattice calculation by Morningstar and Peardon~\cite{Morningstar:1999rf}. In the absence of experimental data one may wonder whether the lattice prediction may be independently tested. As far as the supersymmetric extensions of QCD are concerned even the lattice results are absent due to the technical difficulties one faces extending the lattice methods to fermions.

At strong coupling our main systematic tool -- perturbation theory -- does not apply. String theory offers another systematic tool for the analysis of strongly coupled physics. This tool is based on a weak-strong coupling duality, which is called holographic correspondence. In other words, string theory offers a first principle method of switching from the expansion in $\lambda$, where $\lambda$ is large to an expansion in $1/\lambda$. Essentially the idea is based on the observation of the equivalence of two descriptions: of a quantum gravity (string theory) in a special space and of a field (gauge) theory on the boundary of this space.

In this note we will review the application of the holographic method to the calculation of the glueball spectrum in the $\Nc=1$ supersymmetric Yang-Mills theory. For this purpose the holographic model of Klebanov and Strassler~\cite{KS} will be employed. We make predictions about the masses of low spin glueballs ($0$ or $1$). We also find that the available results, \emph{e.g.} the ratios $m_{1^{+-}}/m_{2^{++}}$ and $m_{1^{--}}/m_{2^{++}}$, are also consistent with the lattice results beyond our expectations.

The review is based on a series of works~\cite{BHM}-\cite{Gordeli:2009nw}, in which the glueball spectrum of the Klebanov-Strassler theory was studied. Some earlier works, see~\cite{Brower:2000rp}, analyze the spectrum in the non-supersymmetric case. In the Klebanov-Strassler setup some issues of the non-supersymmetric holographic calculation are resolved, such as extra degeneracy of the $2^{++}$ and $0^{++}$ states.

\section{Klebanov-Strassler theory}


\paragraph{Gravity solution} The basic model of the holographic correspondence is the duality suggested by Maldacena~\cite{Maldacena:1997re} between the $\mathcal{N}=4$ supersymmetric Yang-Mills theory and type IIB string theory in 10d space $AdS_5\times S^5$. Although by itself the $\mathcal{N}=4$ model is very deep and interesting, \emph{e.g.} at high energies it may qualitatively model QCD, it cannot serve to the purpose of describing the physics of confinement.

As far as confinement is concerned it is also instructive to study $\Nc=1$ theories as their low-energy physics resembles that of QCD. One of the most popular $\Nc=1$ holographic models is the one derived by Klebanov and Strassler (KS)~\cite{KS}. From the string theory point of view KS theory is the type IIB string theory on $AdS_5\times T^{1,1}$, where $T^{1,1}$ is topologically $S^3\times S^2$. In the low-energy limit the string theory becomes the type IIB SUGRA, so that the KS theory is represented by the following solution of the SUGRA equations (see \emph{e.g.}~\cite{BHM}).
\begin{eqnarray}
\text{metric, } \d s^2 & = & h^{-1/2}\eta_{\mu\nu}\d x^\mu \d x^\nu + h^{1/2}\d s_6^2\,,\\
\text{NS 3-form, } H_3 & = & \d B_2 = \d\left(g_s M \,\d\tau\wedge(f\omega_{12} + k\omega_{34} )\right) \,, \\
\text{RR 3-form, } F_3 & = & M\left(g^5\wedge\omega_{34}+\d(F\omega_{13})\right)\,, \\
\text{RR 5-form, } F_5 & = & (1+\ast_{10})B_2\wedge F_3\,,
\end{eqnarray}
while the dilaton $\Phi$ and the RR scalar $C$ can be chosen to vanish. Here $h$ is called the warp-factor\footnote{Due to $h$ the gravity lives in $\mathcal{M}_{1,3}\times \mathcal{C}^6$, where $\mathcal{C}_6$ is a 6d cone with the base $T^{1,1}$.}, $\eta_{\mu\nu}$ is the Minkowski metric, $g_s$ is the string coupling, $M$ is a parameter, encoding the rank of the gauge group in the IR, $\ast_{10}$ is the 10d Hodge operator. The coefficients $h$, $f$, $k$ and $F$ are functions of the $AdS_5$ radial coordinate $\tau$, which is related to the standard $AdS$ coordinate $r \propto \epsilon^{2/3}{\rm e}^{\tau/3}$ (at large $r$). Explicit expressions for these functions and the conifold metric $\d s_6^2$ can be found in~\cite{KS}. $\epsilon$ is the deformation parameter of the conifold -- a dimensionful parameter, related to the strong coupling scale $\Lambda$ in the dual theory.

The above solution is written in terms of the 1- and 2-forms invariant under the $SU(2)\times SU(2)$-symmetry of $T^{1,1}$. These include
\be
\label{basisforms}
g^5\,, \qquad \omega_{12}=g^1\wedge g^2\,, \qquad \omega_{34}=g^3\wedge g^4\,, \qquad \omega_{13}=g^1\wedge g^3 +g^2\wedge g^4\,.
\ee
The definition of the 1-forms $\{g^{i}\}$ is given by equation~(4) in~\cite{KS}.

The geometry of the KS solution exhibits several features reminiscent of QCD or $\Nc=1$ SYM. In the UV limit the solution is almost conformal up to the logarithmic running of the gauge coupling. The geometry has an IR cutoff ($r\sim \epsilon^{2/3}$), which will later define the mass gap. One can also show that it also exhibit the correct pattern of chiral ($U(1)_R$) symmetry breaking.

\paragraph{Dual field theory} Klebanov and Strassler have argued that the above gravity solution describes the following ``cascading'' theory. Let us start in the UV from an $SU(N)\times SU(N+M)$ $\Nc=1$ gauge theory with the matter sector consisting of a pair of doublets $A_{1,2}$ and $B_{1,2}$ in the bifundamental representations of the gauge group:
\be
A_{1,2}\in (N,\overline{N+M})\,,\qquad B_{1,2}\in (\overline{N},{N+M})\,.
\ee
Here $A_i$ and $B_i$ are $\Nc=1$ (chiral) superfields. One can also add a superpotential
\be
\label{superpotential}
W = \lambda \epsilon^{ik} \epsilon^{jl} \Tr A_i B_j A_k B_l\,.
\ee
Notice that the D-term equation in such a theory gives precisely the algebraic definition of the conifold.

As one runs the theory to the IR, the two couplings of the two gauge factors will run in the opposite direction, and the theory will undergo a Seiberg-type duality: $SU(N)\times SU(N+M) \to SU(N)\times SU(N-M)$, to avoid the Landau pole for one of the couplings. One can observe that upon the duality the theory is similar to the original one, with the only difference in the rank of the gauge group. Thus the story repeats itself and one discovers a ``cascade'' of Seiberg dualities, which continues until one of the gauge factors disappears:
\begin{multline}
\label{cascade}
SU(M) \longleftarrow SU(2M)\times SU(M)\longleftarrow \ldots  \longleftarrow \\
\longleftarrow SU(N)\times SU(N-M)\longleftarrow SU(N)\times SU(N+M) \longleftarrow \ldots
\end{multline}
In the last step the theory becomes an $\Nc=1$ $SU(M)$ gauge theory.

In the original paper~\cite{KS} the expectation of Klebanov and Strassler was that the IR fixed point is precisely the $\Nc=1$ SYM theory. However, it was understood later that the IR theory is not precisely the same. The gravity background also realizes spontaneous breaking of the baryon number symmetry, and massless particles appear in the spectrum. As a result some massive mesons made out of $A$ and $B$ fields do not decouple in the IR.

\section{Gravity fields and glueball operators}

\paragraph{Particle spectrum} The spectrum of particles of a theory can be extracted from the poles of the 2-point correlation functions of the field theory operators. Holography provides a prescription for the calculation of any correlator at strong coupling:
\be
\label{correlators}
\langle \mathcal{O}_{i_1}\cdots \mathcal{O}_{i_n}\rangle = \frac{\delta^{(n)} S_{\rm bulk}}{\delta \varphi_{i_1}^{(0)}\cdots \delta \varphi_{i_n}^{(0)}}\bigg|_{\varphi_i^{(0)}=0}\,,
\ee
where $S_\text{bulk}$ is the classical gravity (bulk) action computed on the solutions to the gravity (bulk) field perturbations around the background solution with the following boundary conditions. The asymptotic values of the bulk fields are the sources of the field theory operators:
\be
\label{field-operator}
\delta \varphi_i(r,x^\mu|\varphi^{(0)})  \qquad  \longleftrightarrow \qquad \delta\mathcal{L}=\int\d^4x\, \varphi_i^{(0)} \mathcal{O}_i\,,
\ee
where $\delta\varphi_i$ is a collective notation for the fluctuations of the bulk fields and $\varphi_i^{(0)}$ is their value on the boundary $r\to\infty$.

Boundary condition~(\ref{field-operator}) sets a correspondence between the field theory operators and the gravity (bulk) fields. However the classical gravity approximation corresponds to the low-energy limit, which implies that higher spin operators, with large number of derivatives, are suppressed in the holographic consideration. In particular, only the states with the spin less than 2 can be derived in the lowest holographic approximation, with the exception of the spin 2 state related to the energy-momentum tensor operator.

Correspondence also establishes a relation between the mass of the state in the bulk and dimension of the dual operator. For the scalar field in the bulk
\be
\label{mass-dimension}
\Delta = 2 + \sqrt{4+m_5^2R^2}\,,
\ee
where $m_5$ is the eigenvalue of the Laplace-Beltrami operator on $T^{1,1}$, which equals the mass of the scalar in the reduced 5d equations. For vectors and tensors one can find the mass-dimension relation in \emph{e.g.}~\cite{AdS/CFT2}.

Practically, instead of computing the on-shell action, one finds the linearized bulk equations corresponding to an operator of interest and solve
the Sturm-Liouville problem for the resulting system. That is find the values of the 4d mass of the fluctuation, for which the solution is normalizable.

\paragraph{Quantum numbers} Glueballs are classified by the $J^{PC}$ quantum numbers. Quantum numbers are determined by the representation of the operator, \emph{e.g.}~the classification of~\cite{Jaffe:1985qp}. The quantum numbers of the operators are in turn in a correlation with the quantum numbers of the bulk fields. The spin of the bulk field is determined by the representation of the 4d Lorentz group. It can be shown that any fluctuation of the SUGRA fields can be parameterized in terms of either 0- (scalar), or 1-form (vector). The only possible spin 2 fluctuation is the fluctuation of the metric. This is again a restriction of the gravity limit of the holographic correspondence, mentioned above. Higher spin states are encoded in the string excitations, which are infinitely massive in the low-energy limit.

Parity is a symmetry, which reflects the spatial part of the Minkowski space: $x^i\to - x^i$. The properties of the bulk fields under the parity transformation are determined through the interaction of the bulk fields with probe D3-branes~\cite{Brower:2000rp}. The interaction of the fields on the D3-brane ($A_\mu^a$, $\lambda_\alpha^a$) with the bulk field is given by the DBI action with Chern-Simons terms. The parity of the bulk fields is then fixed from the invariance of the DBI action.

KS theory is not invariant under the usual charge conjugation $C$. However, $C$ supplemented by the exchange $A_i\leftrightarrow B_i$ is a symmetry of the theory. This symmetry was named $\mathcal{I}$-symmetry. Provided this, the $C$-numbers of the bulk fields can be fixed in a similar manner to the parity. The pure gauge $\Nc=1$ sector of the Klebanov and Strassler theory does not contain the $A_i$ and $B_i$ fields. Therefore, for this sector, the $C$ and $\mathcal{I}$ quantum numbers coincide.

\section{Singlet glueballs}

\paragraph{Symmetries} KS theory inherits the $SU(2)\times SU(2)$ isometries of $T^{1,1}$ as its global symmetry. This is reflected in the spectrum of glueballs, which are organized in the representations of this group. Here we will only be interested in the $SU(2)\times SU(2)$-singlet sector of the theory. All the glueball states of the $\Nc=1$ SYM are contained in this sector. For attempts to study the non-singlet sector see \emph{e.g.}~\cite{Pufu:2010ie}.

The glueballs are also organized in the representations of SUSY. We will mostly be dealing with the massive representations (multiplets). These multiplets are characterized by a half-integer number $j$, so that the multiplet consists of the states with the spin $|j-1/2|\oplus j\oplus j\oplus |j+1/2|$. With the exception of the scalar multiplet $j=0$, the multiplets are named by the highest spin state: vector ($j=1/2$), gravitino ($j=1$) and graviton ($j=3/2$). In what follows we will ignore the fermionic components of the multiplets.

\paragraph{Graviton multiplet} The simplest SUSY multiplet to analyze is the graviton multiplet, which contains massive spin 2 and spin 1 fields. The spin 2 state is special as it is produced by the energy-momentum operator $T_{\mu\nu}$, which is present in any theory. This is a $2^{++}$ state. The corresponding gravity fluctuation is the transverse traceless fluctuation of the metric along the Minkowski directions (graviton):
\be
T_{\mu\nu}\qquad \longleftrightarrow \qquad \delta (\d s^2) = h_{\mu\nu}(x^\mu,\tau)\d x^\mu\d x^\nu\,.
\ee

There are no other operators with the same quantum numbers to mix with, correspondingly, there are no other bulk fields to be excited. The linearized equations for the graviton are particularly simple. They give the equation of a scalar coupled to gravity in a minimal way, that is massless Klein-Gordon equation in the KS background. The spectrum of this equation can be approximated by a quadratic fit ($n=1,2,\ldots$)
\be
\label{graviton-spec}
m_n^2 = 0.290 n^2 + 0.528 n + 0.318\, \qquad \text{in units of}\quad {3\epsilon^{4/3}}/({2^{5/3}g_s M\alpha'})\,.
\ee

It is a known result in SUSY theories that energy-momentum operator $T_{\mu\nu}$ enters the same supermultiplet as the current $J_\mu^R$ of the $U(1)_R$ symmetry~\cite{Ferrara:1974pz}. In the Klebanov-Strassler geometry, the $U(1)_R$ symmetry is realized as shifts of one of the angles ($\psi$) on $T^{1,1}$. One can find that this symmetry is anomalous, which has a geometric realization~\cite{Klebanov:2002gr}.

Associated to the $U(1)_R$ is $J_\mu^R$ operator on the gauge theory side. The associated bulk vector excitation ${\bf V}_\mu$ comes from the following ansatz~\cite{Dymarsky:2006hn,Gordeli:2009nw}:
\be
\delta(\d s^2) = {\bf V}_\mu \d x^\mu g^5\,,\qquad \delta F_5 = ({\bf W}\wedge\d g^5 + {\bf W}'\d\tau\wedge g^5)\wedge\d g^5 + \ldots\,
\ee

$U(1)_R$ current has the quantum numbers $1^{++}$ and conformal dimension $\Delta=3$. In fact, there is another $1^{++}$ linear combination of ${\bf V}_\mu$ and ${\bf W}_\mu$~\cite{Gordeli:2009nw}. One can find that the true eigenmodes of the system have $\Delta=3$ and $\Delta=7$, so that one of them is indeed dual to the $U(1)_R$ current, while the other one is dual to a hybrid operator
\be
\mathcal{O}^{1^{++}}_\mu= \frac14\,\Tr\{\lambda,F_{\alpha\beta}\}\sigma^{\alpha}\{\bar{\lambda},{F^\beta}_\mu\}+\Tr\{\lambda,\tilde{F}_{\alpha\beta}\}\sigma^{\alpha}\{\bar{\lambda},{{\tilde{F}}^\beta}_{~\mu}\} + \ldots,
\ee
where ellipses stand for the fermionic terms with derivatives and auxiliary fields.

As expected the spectrum of the dual $U(1)_R$ vector field coincides with the one of the graviton~(\ref{graviton-spec}). In the same units the spectrum of the $\mathcal{O}^{1^{++}}_\mu$ is approximated by the fit
\be
m_n^2 = 0.287 n^2 + 1.80 n + 2.32\,,\qquad \text{for $\mathcal{O}^{1^{++}}_\mu$} \,.
\ee

\paragraph{C-odd sector} The next in difficulty is the sector with $\mathcal{I}$ odd. First, this sector contains two massless scalars ($P=\pm 1$). The existence of these states was demonstrated in~\cite{Gubser:2004qj}, where it was argued that the pseudoscalar is a Goldstone mode of the baryon number symmetry $U(1)_B$, spontaneously broken by the expectation values of the baryonic operators produced in the last step of the cascade~(\ref{cascade}). The dual operator of the pseudoscalar is the $\Delta=4$ operator:
\be
\label{BaryonCurrent}
\partial_\mu {J^B}^\mu = {\rm Im}\,\Tr(a_i^\ast \Box a_i - b_i^\ast\Box b_i ) + \, \text{fermionic terms}\,,
\ee
while the scalar operator is the real part of the same expression. Here $a_i$ and $b_i$ are the scalar components of the superfields $A_i$ and $B_i$. As expected, the two massless states form one $CP$-extended scalar multiplet and can be described by a single complex operator.

The operator $J^B_\mu$ and its superpartners can also produce massive states. Consider all possible $SU(2)\times SU(2)$-singlet (pseudo-) scalar fluctuations~\cite{Benna:2007mb}:
\begin{eqnarray}
\label{scalar-}
\delta B_2 & = & \chi\, \d g^5 + \partial_\mu\sigma\, \d x^\mu\cdot g^5\,, \qquad \delta g_{13} = \delta g_{24}= \psi\,; \\
\label{pseudoscalar-}
\delta C_2 & = & \tilde{\chi}\, \d g^5 + \partial_\mu\tilde{\sigma}\, \d x^\mu\cdot g^5\,,
\end{eqnarray}
where the metric is excited along the $SU(2)\times SU(2)$ invariant direction specified by the interior product $g^1\cdot g^3 + g^2\cdot g^4$. Notice that (\ref{scalar-}) describe a scalar, while (\ref{pseudoscalar-}) -- a pseudoscalar excitation.

Analysis of the linearized equations shows that only two of the functions $\chi$, $\psi$ and $\sigma$ are independent. As a result there are two mixed excitations producing the $0^{+-}$ eigenstates with the dimensions of the dual operators $\Delta=2$ and $\Delta = 5$. Analysis of the superconformal representations~\cite{Ceresole:1999zs} shows that
\be
\label{operator0+-}
\mathcal{O}^{0^{+-}}_2= \Tr(a_i^\ast a_i - b_i^\ast b_i )\,,\qquad \mathcal{O}^{0^{+-}}_5= \frac{1}{2}\,{\rm Re}\,\Tr D\{\lambda,\lambda\}\,,
\ee
where $D$ is the auxiliary field of the pure gauge $\Nc=1$ sector.

For the pseudoscalar fluctuation~(\ref{pseudoscalar-}) the functions one finds only one pseudoscalar. The $0^{--}$ state will be given by the imaginary part of the above complex operator $\Tr D\{\lambda,\lambda\}$.

Numerical analysis gives the following for the spectrum~\cite{Benna:2007mb} in units of~(\ref{graviton-spec}):
\begin{eqnarray}
 m_n^2 & = & 0.262 n^2 + 0.130 n + 0.340\, \qquad \text{for $\mathcal{O}_2^{0^{+-}}$}\,, \\
 m_n^2 & = & 0.277 n^2 + 1.79 n + 2.17\, \qquad \text{for $\mathcal{O}_5^{0^{+-}}$}\,,\\
 m_n^2 & = & 0.289 n^2 + 1.15 n + 0.996\, \qquad \text{for $\mathcal{O}^{0^{--}}$}\,,
\end{eqnarray}

The $\mathcal{I}$-odd sector also contains 7 vectors, 4 of which belong to the parity even sector and 3 to parity odd~\cite{Dymarsky:2008wd}. $P$-even states can be derived from the following bulk fluctuations, written in terms of the $SU(2)\times SU(2)$-invariant forms on $T^{1,1}$,
\begin{eqnarray}
\delta B_2 & = & {\bf J}\wedge \d\tau\,, \qquad \delta C_2 =  \ast_4 \d_4 {\bf D} + {\bf C}\wedge g^5\,, \\
\delta F_5 & = & (1+\ast_{10})\big({\bf F}\wedge\d\tau\wedge \omega_{12}\wedge g^5 + {\bf G}\wedge\d\tau\wedge \omega_{34}\wedge g^5 + \\
& + & ({\d_4 {\bf P}}\wedge \omega_{12} + {\d_4 {\bf Q}}\wedge \omega_{34})\wedge g^5 + \d_4{\bf R}\wedge \d\tau\wedge\omega_{13}\big)\,,
\end{eqnarray}
where $\ast_4$ denotes the 4d Hodge operator, $\d_4$ is the exterior derivative acting in Minkowski space. Bold face is used to denote 1-forms.

Linearized equations can be partially diagonalized analytically separating two eigenvectors and a system of 2 coupled equations. The eigenvalues of the latter can be found numerically and correspond to the $1^{+-}$ states dual to the $\Delta=3$ $J_\mu^B$ operator and $\Delta=6$ superpartner of $\mathcal{O}_5^{0^{+-}}$. Two remaining eigenvectors belong to two "gravitino" multiplets, which contain parity even and odd components of $\Delta=5$ and $\Delta=6$ operators:
\be
\mathcal{O}^{(5)}_{\mu} = \Tr F_{\mu\nu}\lambda\sigma^\nu\bar\lambda + \ldots \,,\qquad \mathcal{O}^{(6)}_{\mu\nu}= \Tr F_{\mu\nu}F_{\rho\sigma}F^{\rho\sigma}+\ldots \,,
\ee
where ellipses stand for higher order fermionic terms and terms with auxiliary fields. Numerical calculation of the spectrum of the gravitino multiplets gives~\cite{Dymarsky:2008wd}
\be
 m_n^2  =  0.287 n^2 + 1.02 n + 0.633\,, \quad 
 m_n^2 =  0.288 n^2 + 1.31 n + 1.44\,. 
\ee


\paragraph{Scalar multiplets} The $\mathcal{I}$-even scalar sector is the most difficult one to study. The reason is a heavy mixing among different states. The original study of the spectrum of the $0^{++}$ states was performed in~\cite{BHM}. It was shown that there are 7 independent excitations of the bulk fields, which can be parameterized according to table~\ref{tab scalars}.

\begin{table}[htb]
\begin{center}
\begin{tabular}[c]{|c|c|c|c|}\hline \Trule\Brule
Mode  & 10d Fluctuation & $\Delta$ & $R$ \\
\hline
\Trule $\Phi$ &  $\delta\Phi$ &  4 & 0 \\
$s$ &  $\delta B_2$: $\omega_2 \propto g^1\wedge g^2 + g^3\wedge g^4$  & 4 & 0 \\
$y$ &  $\delta \d s^2$: $(g^1)^2+ (g^2)^2 -(g^3)^2-(g^4)^2$  &  3 & $\pm 2$ \\
$N_2$ &  $\delta (C_2+iB_2)$ &  3 & $2$ \\
$f$ &  $\delta \d s^2$: $-\frac{10}{9}\,(g^5)^2 + 2\d s_{T^{1,1}}^2$  & 6 & $0$ \\
$N_1$ &  $\delta (C_2+iB_2)$  &  7 & $- 2$ \\
\Brule $q$ &  $\delta \d s^2$: $-5\d s_5^2 + 3\d s_{T^{1,1}}^2$ & 8 & $0$ \\
\hline

\end{tabular}
\end{center}
\caption{\small $\mathcal{I}$-even scalar fluctuations of the KS background found in~\cite{BHM}. The dimension $\Delta$ of the dual operator and the $R$-charge of the mode shown.}
\label{tab scalars}
\end{table}

The system of 7 linearized equations obtained in~\cite{BHM} is strongly entangled. A tremendous breakthrough of~\cite{BHM} was the computation of the collective spectrum of all 7 $0^{++}$ particles. Naive attempts to identify any single tower of states in the collective spectrum may lead to wrong results. One may observe from the study of other sectors, that the spectrum of any individual tower is excellently fit by a quadratic formula. but for the $0^{++}$ spectra, one would always fail to fit the lightest states. Perhaps, there is a deviation from the quadratic dependence at least for some eigenstates.

In~\cite{Gordeli:2009nw} it was shown how SUSY can help to resolve the problems (at least partially). It turns out that one of the $1^{++}$ states is degenerate with one of the $0^{++}$ states. Together they form a massive vector supermultiplet. Thus, one out of 7 towers can be extracted from the collective spectrum. The question is whether the remaining 6 towers can be disentangled. This question was addressed in~\cite{Gordeli:20xx}, where the pseudoscalar $0^{-+}$ fluctuations of the metric were investigated. It was shown that the pseudoscalar equations describe 6 independent modes necessary to complete 6 massive scalar multiplets with the remaining $0^{++}$. The spectrum of the system is to be compared with~\cite{BHM}.

\section{Summary}

We have summarized the results of the calculation of the glueball spectrum in the $SU(2)\times SU(2)$-singlet sector of the KS theory. Since one of our motivations was a prediction of the spectrum of the pure glue $\Nc=1$ supersymmetric Yang-Mills theory, let us discuss the result from this perspective. We remind that the gravity approximation only allows to compute the spectrum of low spin states.

\begin{figure}[htb]
 \centering
\begin{minipage}{0.45\linewidth}
  \includegraphics[width=\linewidth]{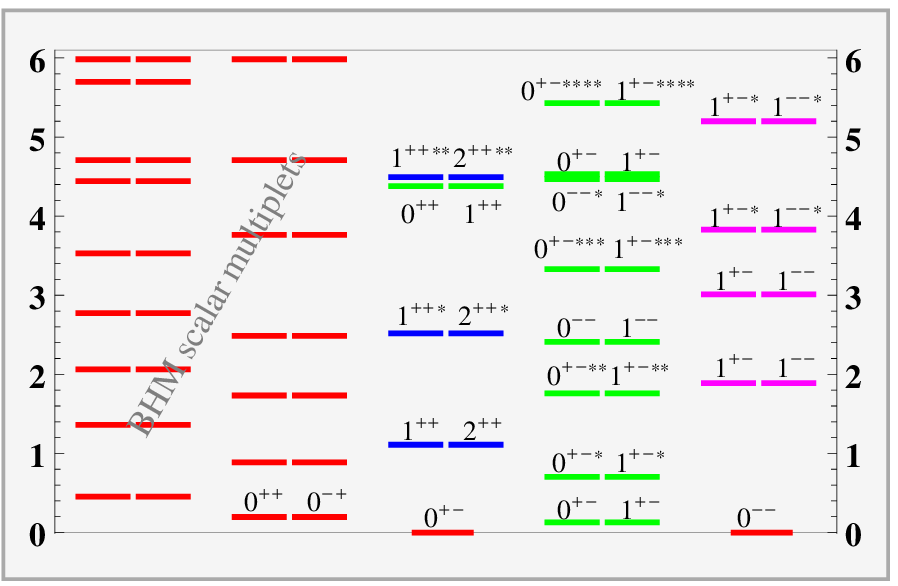}
 \end{minipage}
\hfill{
\begin{minipage}{0.47\linewidth}
 \includegraphics[width=\linewidth]{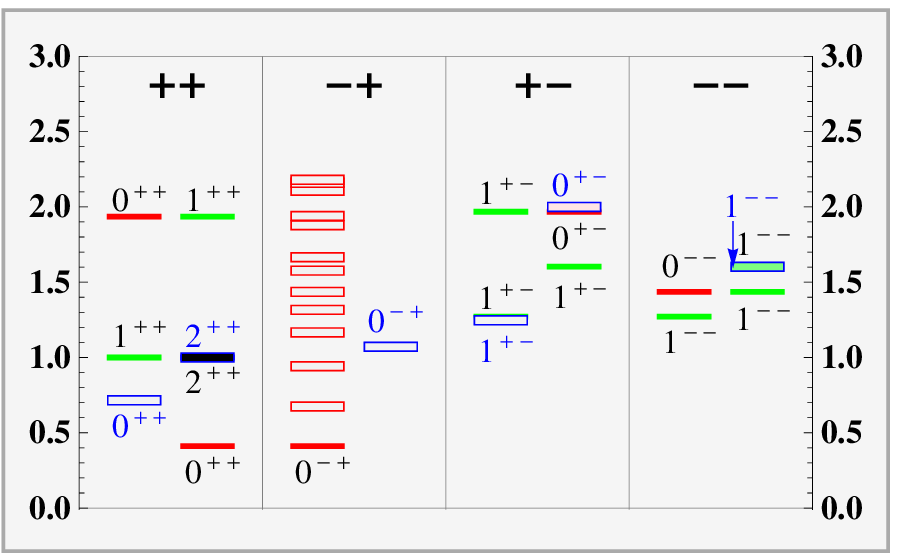}
\end{minipage}
}
 \caption{Left: $SU(2)\times SU(2)$-singlet (bosonic) spectrum ($m^2$) of the Klebanov-Strassler theory from holography in units~(\ref{graviton-spec}). 2 columns on the left show the collective spectrum of 6 towers of the scalar multiplets. Right: Conjectured (bosonic) spectrum ($m$) of the $\Nc=1$ SYM in units of the $2^{++}$ mass. Only the states with spin $\leq1$ and the $2^{++}$ can be computed in the gravity approximation. The collective spectrum of 6 towers of the scalar multiplets is shown in the $-+$ box (empty red boxes, no labels). Lattice prediction for certain states are shown (empty blue boxes, blue labels).}
 \label{fig:totalspec}
\end{figure}

The total spectrum of the singlet sector is shown on figure~\ref{fig:totalspec}(left). To obtain the spectrum of the $\Nc=1$ SYM theory, we need to throw away the states that contain the $A_i$ and $B_i$ fields. The adjusted spectrum is presented in figure~\ref{fig:totalspec}(right), where we have left only the lightest states from each multiplet tower (with the exception of the scalar towers, for which we cannot identify the lightest states). All the masses are given in units of the $2^{++}$ mass. We have also included the positions of the glueballs in the pure glue non-supersymmetric $SU(3)$ Yang-Mills theory from the lattice calculation~\cite{Morningstar:1999rf}.

Comparing with the lattice results, one can see a nice agreement for the $1^{+-}$ and $1^{--}$ states. The result for the $0^{+-}$ state looks tantalizing, but one should bear in mind that in the SUSY theory the $0^{+-}$ contains fermions~(\ref{operator0+-}), while in the non-SUSY case the underlying operator contains higher derivatives and should not be visible in the gravity approximation.

The scalar sector of the spectrum is still under investigation. Although the mass eigenvalues are known, there is still a large ambiguity in the assignment of the operators to the eigenmodes. There are 6 scalar supermultiplets for which only collective spectrum is shown. Hopefully SUSY will help to identify individual towers from the collective spectrum. It would be interesting to compare the lattice result for the $0^{++}$ and $0^{-+}$ states with the holographic prediction.

\paragraph{Acknowledgements} We would like to thank A.~Dymarsky for useful conversations. DM would like to thank the hospitality of the Particle Physics group of the Weizmann Institute of Science. This work was supported by MCTI and CAPES Foundation -- Brazil, the IIP-MIT exchange program, the RFBR grant \#12-01-00525, the grant of the Russian Ministry of Education and Science \#8207 and the grant of the President of the Russian Federation for Support of Scientific Schools NSh-3349.2012.2.

\end{document}